\documentclass[conference]{IEEEtran}
\IEEEoverridecommandlockouts
\usepackage{cite}
\usepackage{amsmath,amssymb,amsfonts}
\usepackage{listings}
\usepackage{algorithmic}
\usepackage{graphicx}
\usepackage{textcomp}
\usepackage{xcolor}
\usepackage{subcaption}
\usepackage{hyperref}

\colorlet{punct}{red!60!black}
\definecolor{background}{HTML}{EEEEEE}
\definecolor{delim}{RGB}{20,105,176}
\colorlet{numb}{magenta!60!black}

\lstdefinelanguage{json}{
    basicstyle=\small\ttfamily,
    numberstyle=\scriptsize,
    stepnumber=1,
    numbersep=8pt,
    showstringspaces=false,
    breaklines=true,
    columns=fullflexible,
    backgroundcolor=\color{background},
}

\def\BibTeX{{\rm B\kern-.05em{\sc i\kern-.025em b}\kern-.08em
    T\kern-.1667em\lower.7ex\hbox{E}\kern-.125emX}}
\begin{document}

\title{Evaluating Serverless Architecture for \\
 Big Data Enterprise Applications}

\author{\IEEEauthorblockN{Aimer Bhat}
\IEEEauthorblockA{University of Illinois \\
aimerb2@illinois.edu}
\and
\IEEEauthorblockN{Heeki Park}
\IEEEauthorblockA{University of Illinois \\
heekip2@illinois.edu}
\and
\IEEEauthorblockN{Madhumonti Roy}
\IEEEauthorblockA{University of Illinois \\
roy29@illinois.edu}
}

\maketitle

\begin{abstract}
Migration of enterprise applications to the cloud has been driven by a myriad of benefits ranging from availability of infinite computing resources to the elimination of upfront CapEx cost. However, many users still face the burden of complex framework knowledge requirement to efficiently deploy applications in the cloud. In this paper,  we investigate serverless computing for performing large scale data processing with cloud-native primitives. Serverless computing environment abstracts all infrastructure handling, simplifying developers' work who aspire to deploy applications on the cloud. With dynamic input load on the system, serverless architecture has promise to provide better resource utilization and  lower costs.
\end{abstract}


\section{Introduction}
The ecosystem has evolved from enterprise applications developed with conductive data processing using custom-built applications, data warehouse systems, to more recently MapReduce frameworks. Google released its MapReduce paper in 2004 \cite{b1}, and Yahoo open sourced its Hadoop software in 2008 \cite{b2}, thus democratizing big data processing, enabling both startups and large enterprises to take advantage of this new paradigm. This became the foundation of an ecosystem of projects that arose to satisfy the growing needs and requirements of this field. While these frameworks do simplify the life of a data engineer or data scientist, they brought with them the complexity of managing the underlying infrastructure and the learning curve to develop and deploy on these new platforms. Because of this, companies like Hortonworks and Cloudera emerged to help enterprises manage on-premise installation of their growing Hadoop infrastructure. Likewise in cloud, Eric et al \cite{b3} notes cloud computing relieved users of physical infrastructure management but left them with a proliferation of virtual resources to manage. Cloud providers like Amazon Web Services (AWS) developed offerings like the Amazon Elastic MapReduce (EMR) service, which would allow customers to deploy packaged versions of Hadoop software that are pre-tested and validated. However, even with these offerings, an architect or developer who is new to these platforms and this way of thinking, will face a fairly steep learning curve to get started \cite{b4}. Bringing up an enterprise-ready platform with custom integrations is non-trivial and learning to express data processing pipelines in these paradigms takes time to do so efficiently and effectively. The cost model for serverless infrastructure is true pay-per-use, since no costs are involved for the user unless the function is invoked, as opposed to traditional cost models for other cloud offerings, where Virtual Machines are typically billed per-second of uptime regardless of their actual usage \cite{b5}.

We explore serverless computing environments for performing MapReduce processing. In this paper, we discuss our experimentation, challenges faced and learnings based on deploying a data processing workflow on AWS  Serverless infrastructure.

\section{Literature Survey}

Traditional batch processing systems build upon predictability at deployment time for resources needed for data processing. Only small allocation adjustments are made to cope up with unexpected events like system failures \cite{b6}. New solutions like EMR have resource elasticity managed at the virtual machine level, which improves upon the fixed cluster deployments but still has room for improvement \cite{b7}.  Serverless computing platforms integrate support for scalability, availability, load balancing, monitoring, migration and fault tolerance capabilities directly as features of the framework. It provides faster startup times, lower cost and individual task based scaling \cite{b8}. Serverless helps enterprises to focus on application rather than infrastructure \cite{b5}. In this paper, we explore usage of serverless infrastructure for adaptive provisioning of batch processing workload to optimize system utilization.

Since its introduction, serverless offerings from hyperscale cloud service providers have grown from supporting light use-cases, to supporting a wide variety of workloads \cite{b8} \cite{b9} including large scale computations \cite{b10}. Serverless includes Function as a Service (FaaS) offerings like AWS Lambda and other services like Amazon S3 (large object storage), Amazon DynamoDB (key-value storage), Amazon SQS (queuing services), Amazon SNS (notification services), and more \cite{b11}. This entire infrastructure is managed and operated by AWS. Lynn reviews all these platforms and identifies AWS Lambda as the de facto base platform for enterprise serverless cloud computing \cite{b12}. Serverless architecture is evaluated as a paradigm for big data processing workloads in  \cite{b13} \cite{b4} \cite{b14} \cite{b15} \cite{b16}. Sebastian shows how serverless deployment can reduce TCO without compromising on system quality metrics \cite{b14}.

Based on the literature survey of related work in serverless MapReduce, we identify notable work models. Lambda based implementation of MapReduce is documented in an AWS reference architecture blog post \cite{b17}. The document does not detail the tradeoff of using serverless options as compared to Apache Spark based workflow. We also reviewed open-source platforms like PyWren \cite{b18}, Ooso \cite{b19} and Corral \cite{b20} which facilitate big data workflows on serverless architecture. As compared to previous work, our paper focuses on evaluating usage of serverless architecture for MapReduce with highly varying user load behavior and specific targets for cost and response time. There are works like \cite{b21}\cite{b22} which evaluate serverless architecture as a platform for any application but not focussed around data pipelines.

Lloyd notes that there is a lack of repeatable empirical research on the design, development, and evaluation of enterprise applications for serverless architecture \cite{b7}. Tradeoffs related to cost and performance are hence unexplored in this space. As noted by Sill in Cloud computing magazine article \cite{b23}, due to lack of insight into serverless architecture and design, adoption of serverless computing in enterprise applications transitioning to cloud is low. 

We found considerable work around improving elasticity for VM instances \cite{b24} \cite{b25}. They range from machine learning techniques to support scaling of VMs to predicting future demands for autoscaling. As part of this paper, we will delve deeper into factors that impact elasticity in serverless compute platforms. To get optimal cost, serverless scales with concurrent executions and the operator does not need to worry about provisioning or deprovisioning instances per se, as this is fully managed by the cloud provider \cite{b26}.

\section{MapReduce Model}
In big data processing workflows, the data typically enters a number of general phases: ingestion, cleansing, computation, and aggregation \cite{b27}. The ingestion phase is where the raw data is captured in some storage medium, typically some type of distributed file system, e.g. HDFS, S3, or some streaming system, e.g. Kafka, Kinesis. The cleansing phase is where the raw data is massaged into some canonical form that is useful for downstream computation and can sometimes comprise multiple phases of cleansing and transformation. The computation phase is where business logic is applied to the canonical data and prepared for downstream aggregation. The aggregation phase computes summary statistics on the data from the computation phase.

We aim to investigate the use of native AWS services to create a framework for ingesting raw data, performing pre-processing of the raw data into a canonical form, emitting the data as micro-batch messages, computing business logic against those micro-batches, and summarizing the data resulting from the micro-batch computation. We believe serverless style of computing will provide significant cost benefit to enterprises who anticipate fixed compute processes but with dynamic input sizes. Concretely, data flow jobs that have significant variance in execution time and periodicity may result in significantly under-utilized compute infrastructure. Meanwhile, by leveraging cloud-native and serverless services on AWS, we use only the compute resources that are necessary for the workload with no expenses incurred on idle compute infrastructure.

As a proof of concept, we aim to perform analysis on the Bureau of Transportation Statistics data set \cite{b28}, specifically the Airline On-time statistics data set to determine the viability of such a computing model. This data set we used has a total of 240 CSV files, containing monthly data (including airline codes, origin and destination, scheduled and actual arrival and departure times for flights, etc.) reported by US certified air carriers from 1988-2008. Our MapReduce model will entail the ingestion of these CSV files into S3, the use of AWS Lambda (FaaS) functions \cite{b29} for all compute processing, the use of S3 and DynamoDB for all shuffle storage processing, and the use of DynamoDB for storing summary outputs.

As we build out this model, we aim to analyze the performance characteristics of the solution. We also aim to document cases where this solution makes sense and where it does not. We also aim to outline the limitations of the solution and additional work required to make this a solution that can be more broadly adopted.

\section{Prototype Architecture}
\begin{figure}[htbp]
\centerline{\includegraphics[width=\textwidth,height=\textheight,keepaspectratio]{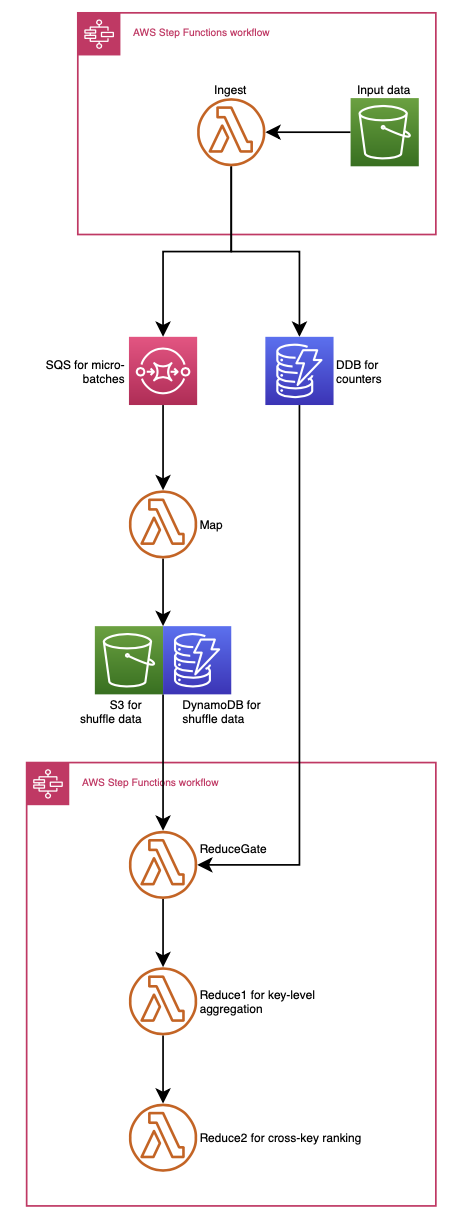}}
\caption{Architecture Diagram}
\label{architecture}
\end{figure}

The current implementation includes three core phases (ingestion, map, reduce) and leverages the following AWS services: AWS Lambda, Amazon SQS, Amazon DynamoDB, and Amazon S3. Our implementation follows the same paradigm in the MapReduce framework by having a scale-out map phase and an aggregation reduce phase with an intermediate shuffle phase. The intermediate shuffle phase is important as we want to ensure that all map operations are completed prior to beginning the reduce operations. In our implementation, we execute a query against the data set for identifying the top 10 airlines by on-time arrival performance. A reference SQL query to get this result via Spark is shown below:

\begin{lstlisting}[linewidth=\columnwidth,language=SQL,breaklines=true,firstnumber=1]
SELECT UniqueCarrier, sum(ArrDelay)/count(*) as OnTimePerformance
FROM df
GROUP BY UniqueCarrier 
ORDER BY 2 ASC
\end{lstlisting}

Our goal is to implement a similar query via Serverless MapReduce framework. The prototype architecture is depicted in Figure~\ref{architecture}.

\subsubsection{Ingestion}
The goal of the ingestion phase is to shard the input data, akin to sharded data in HDFS, slicing up the source data into smaller micro-batches for downstream processing in the map phase. The workflow begins with uploading CSV files into an S3 bucket, which emits a PutObject event, which then triggers the invocation of an AWS Lambda function. The ingest function receives as input the metadata about the object, which includes the bucket and object key. The function then effectively downloads that object and processes the body of that CSV file, which entails the creation of micro-batches of configurable size (default=100). Each of those micro-batches are then emitted to an SQS queue for downstream ordered processing. Each time a message is emitted to the queue, an atomic counter is incremented in DynamoDB to keep track of the number of records that have been ingested.

\subsubsection{Map}
The goal of the map phase is to take the sharded data and to perform the necessary business logic in preparation for downstream aggregation or summarization. Here the workflow continues with a function invocation, triggered by the enqueued messages from the ingestion phase. The function receives as input the micro-batch data, which we filter based on a query parameter criteria. The data that passes the filter criteria is then stored in an S3 bucket as a JSON object or stored in DynamoDB as an item. When processing on the microbatch is completed, a second atomic counter is incremented in DynamoDB to keep track of the number of map tasks that have been successfully processed.

\subsubsection{Shuffle}
The goal of the shuffle phase is to ensure that all ingested micro-batches have been successfully processed in the map phase. Concretely, each job will have an associated execution id, ingested count, and mapped count, as below:

\begin{lstlisting}[language=json,firstnumber=1]
{"id": {
  "S": "a3c9e821-8260-4950-806d-65f7d2e8e989"
  },
  "ingested": {
    "N": "12"
  },
  "mapped": {
    "N": "12"
  }
}
\end{lstlisting}

A gate is introduced in front of the reduce phase, which checks for equivalence of the ingested count and the mapped count. If those values are equal, the gate allows execution to proceed to the reduce phase. If those values are not equal, the gate will sleep for 1 second and perform the check again.

\subsubsection{Reduce}
The goal of the reduce phase is to perform the necessary aggregation computation. Here the workflow continues with a function invocation, that then reads the shuffle data that was outputted from the map phase. In our example, we perform an S3 bucket scan along a prefix that matches our execution id to read the filtered data or a DynamoDB query on a local secondary index with the execution id as the hash key and the airline id as the sort key and then perform the summary statistics .

\begin{table*}[t]
\caption{Execution time}
\begin{center}
\begin{tabular}{cccccc}
\textbf{Function}&\textbf{Total Count} & \textbf{Init Count} & \textbf{Avg Init (ms)} & \textbf{Avg Duration (ms)} & \textbf{\% Init} \\
\hline
ingest & 78 & 35 & 853 & 59110 & 44.87\\
map & 560 & 34 & 811 & 2719 & 6.07 \\
reduce1 & 314 & 125 & 823 & 20606 & 39.81\\
reduce2 & 21 & 11 & 860 & 61 & 52.38\\
\hline
\end{tabular}
\label{table:overall}
\end{center}
\end{table*}

\begin{table*}[t]
\caption{Configuration parameters}
\begin{center}
\begin{tabular}{cccccccc}
\textbf{Scenario}&\textbf{ShuffleSystem} & \textbf{Files} & \textbf{Threads} & \textbf{IngestMemMB} & \textbf{MapMemMB} & \textbf{Reduce1MemMB} & \textbf{Reduce2MemMB}\\
\hline
1 & S3 & 1 & 1 & 2048 & 128 & 10240 & 128\\
2 & S3 & 1 & 2 & 2048 & 128 & 10240 & 128 \\
3 & S3 & 1 & 3 & 3072 & 128 & 10240 & 128\\
4 & S3 & 1 & 3 & 3072 & 1024 & 10240 & 128\\
5 & S3 & 12 & 3 & 3072 & 1024 & 10240 & 128\\
6 & DynamoDB & 12 & 3 & 3072 & 1024 & 10240 & 128\\
\hline
\end{tabular}
\label{table:params}
\end{center}
\end{table*}

\section{Experimentation Setup}

\subsection{Metrics overview}
We use several Key Performance Indicators (KPIs) in order to identify best serverless instance configuration and end-to-end MapReduce execution optimizations:
\begin{itemize}
\item Invocation requests - Number of incoming requests to execute a Lambda function
\item Invocation duration - Execution duration of a given invocation request
\item Concurrency - Invocation requests * Invocation duration
\textit{Note that concurrency is the number of requests that your function is serving at any given time}\cite{b37}
\item Initialization duration - When the Lambda service allocates an instance of the function, it loads the function runtime and executes the initialization code, which happens once per new instance of a function \cite{b38}
\end{itemize}

\subsection{Configuration overview}
Some data points sampled from executions of the overall flow are shown in Table \ref{table:overall}. This execution data was also used to initialize the memory size of the Lambda functions.

Configuration parameters of a number of tuned performance tests run against theMapReduce system are listed in Table \ref{table:params}.

\begin{figure}[h]
\centerline{\includegraphics[width=0.57\linewidth]{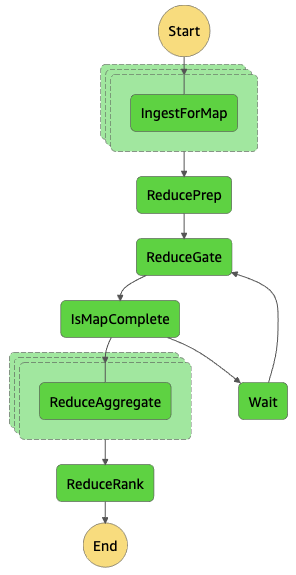}}
\caption{Overall architecture}
\label{archflow}
\end{figure}

\section{Architecture overview} \label{archoverview}
The implementation for the performance data gathering phase focused on a single query based on the airline data, e.g. top 10 airlines by on-time arrival performance. The implementation was conducted and measured in three distinct phases: ingest, map, reduce, as shown in Fig.~\ref{architecture}.

Fig.\ref{archflow} is a visualization of the overall flow, which is implemented as an AWS Step Function standard workflow. For every job execution, we generate an execution id, which is passed along each phase of the execution, so that we can track the activity through each discrete component, similar to a correlation or trace id in distribution application monitoring. Note that the Map function specifically is not present in the Step Function workflow as it triggers directly off of the SQS queue, rather than being orchestrated by the Step Function workflow. This is the reason we have a reduce gate with counters at the ingest and map phase, to ensure that all map work is completed prior to proceeding to the reduce phase.

\subsection{Ingest}

For the ingest phase, all 240 raw data files (34,870MB)  were uploaded to an S3 bucket. In a production scenario, an S3 event trigger can be configured to invoke the ingest function upon an s3:ObjectCreated event \cite{b36}. However, for the purposes of performance tests, an AWS Step Function standard workflow was configured to invoke the ingest function in parallel with each ingest Lambda function processing one of the raw data files. 

A Python script was written to create an event payload by listing the contents of the S3 bucket prefix, which was subsequently used to asynchronously execute the Step Functions workflow. This workflow iterates through the list of raw files that need to be processed. Each of those files is passed as an input parameter to the ingest Lambda function, which then processes the elements of that raw file.

\begin{figure}[h]
\centerline{\includegraphics[width=0.9\linewidth]{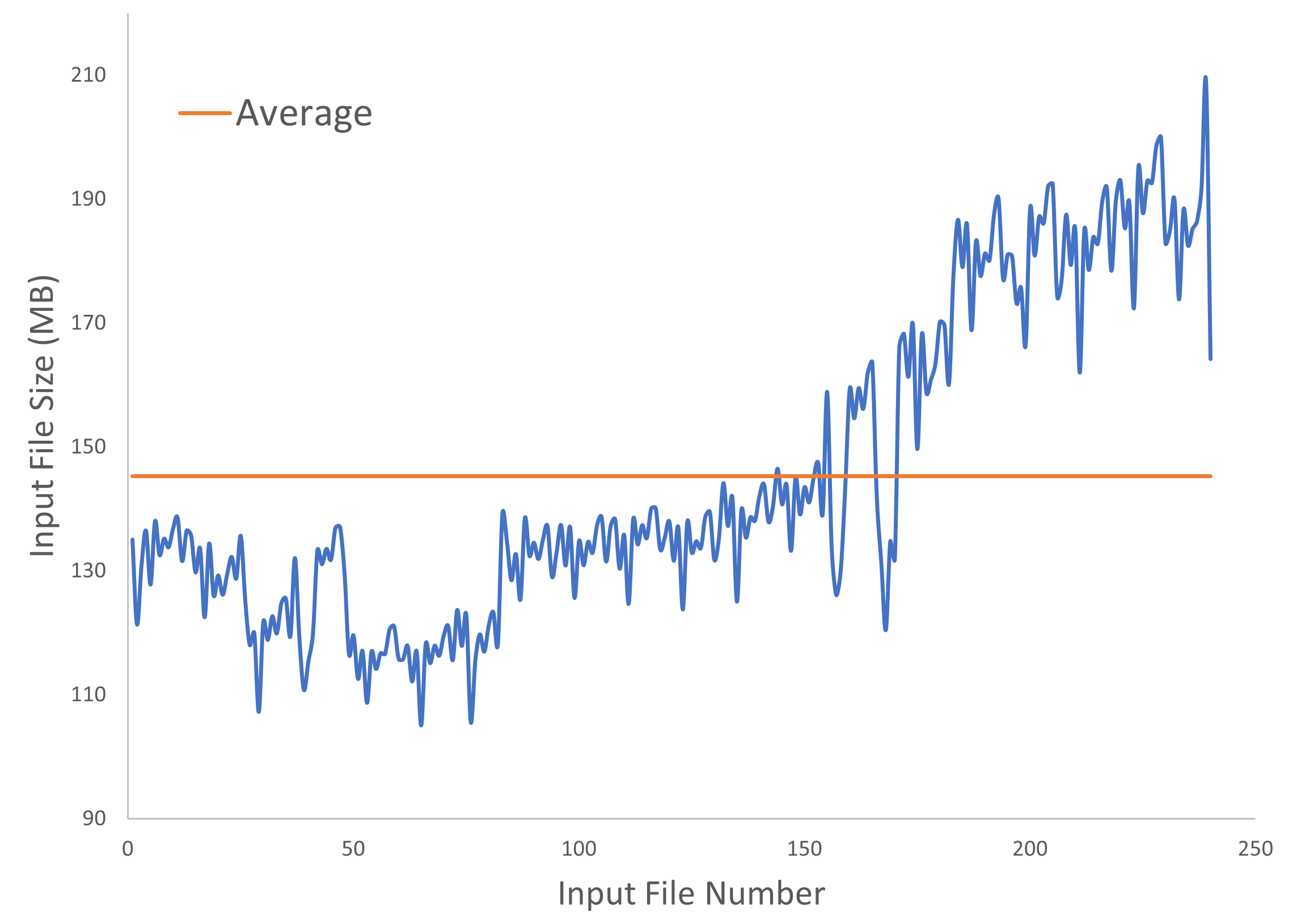}}
\caption{Input file size}
\label{filesize}
\end{figure}

The average file size was 145.2MB as shown in Fig.\ref{filesize} and the average processing time for the ingest function configured with 2GB of memory was 88.4 seconds.

\begin{table*}[t]
\caption{Lambda usage}
\begin{center}
\begin{tabular}{ccccccc}
\textbf{Processed}&\textbf{LambdaMemMB} & \textbf{DurationMs} & \textbf{MaxMemUsedMB} & \textbf{InitDurMs} & \textbf{CapMB}  & \textbf{TotalMB} \\
\hline
436950 & 1024 & 121133.14 & 415 & 500.28 & 135.0045462 & 34869.82844 \\
436950 & 2048 & 88402.71 & 415 & 489.43 & 135.0045462 & 34869.82844 \\
436950 & 4096 & 90243.8 & 415 & 518.14 & 135.0045462 & 34869.82844 \\
\hline
\end{tabular}
\label{table:lambda}
\end{center}
\end{table*}

\subsubsection*{Optimal Lambda function}
We performed testing of the ingest function at 1GB, 2GB, and 3GB to determine the optimum memory when trading off cost v/s performance. Lambda allocates CPU and other resources proportional to memory configuration. We estimate that at the CPU levels specified above, we had approximately 2, 2, and 3 vCPU available for processing, and thus configured parallel processes for the ingest function accordingly. CPU ceilings may be slightly different, according to some public research \cite{b41}.

Results shown in Table~\ref{table:lambda} were gathered by multiple invocations of the function. The 2GB lambda function gives us a significant speed up which helps keep the cost same as 1GB lambda function.  Usage of the power tuning tool also provided us with a similar workpoint.

When attempting to parallelize processing of the raw data, our initial implementation used the asyncio libraries, which were subjected to a single thread due to the global interpreter lock limitation within Python. We then switched to using the concurrent.futures.Executor library, which allowed us to run multiple processes in parallel. We configured the executor workers to match the number of anticipated vCPU, e.g. 1GB memory with 2 vCPU with 1 worker, 2GB memory with 2 vCPU with 2 workers, 3GB memory with 3 vCPU with 3 workers. By doing this, we saw execution time for the ingest function, which was processing one 135MB file average 92.2 seconds, 63.1 seconds, and 54.0 seconds respectively. We observed that we were experiencing diminishing returns on increasing the parallelism on the ingest function and thus left it at 3GB. At this point, we observed that the map function was beginning to bottleneck the overall flow, which we will cover in the next section.


The ingestion of 1, 12 and 60 file(s) resulted in 436,950, 5,202,096 and 25,683,271 records being processed through this pipeline respectively.

\begin{figure*}[h]
\centering
\begin{subfigure}{0.49\textwidth}
\centering
\includegraphics[width=\linewidth]{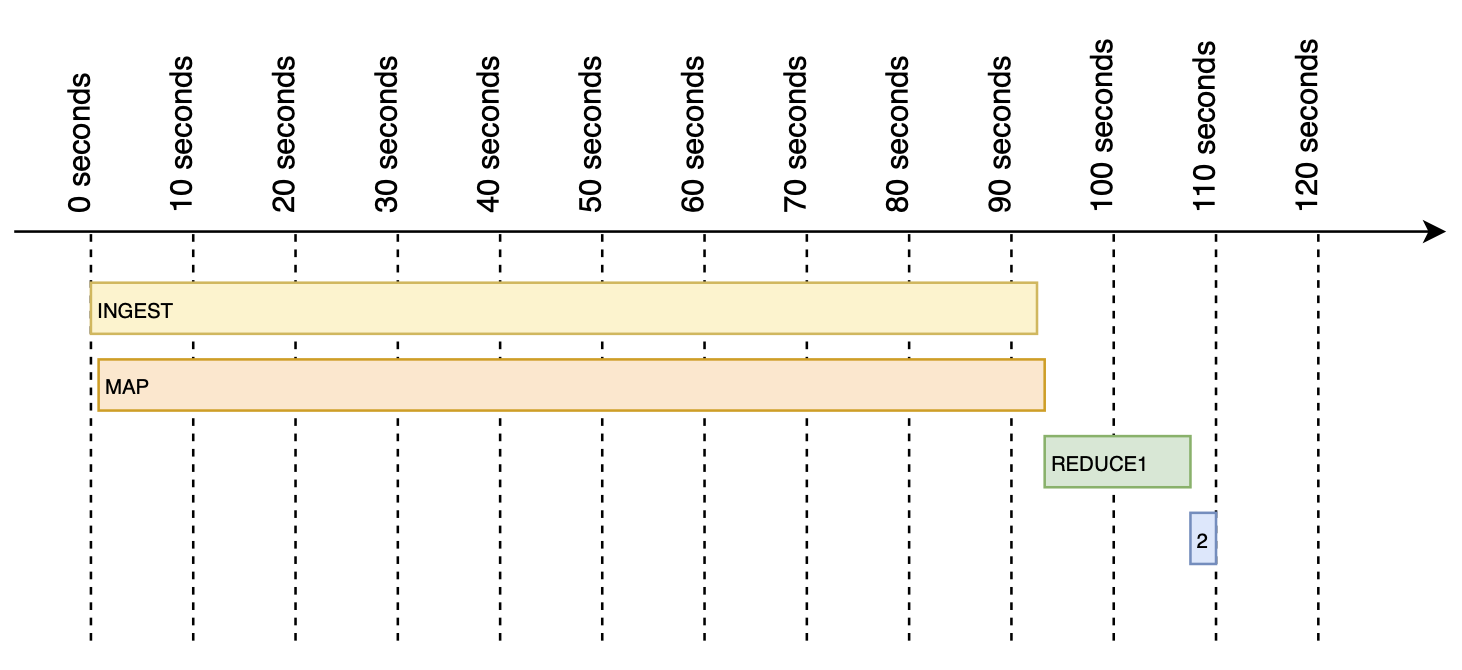}
\caption{1 thread for ingest, 2048/128/10240/128MB lambda memory}
\label{sc1}
\end{subfigure}
\begin{subfigure}{0.49\textwidth}
\centering
\includegraphics[width=\linewidth]{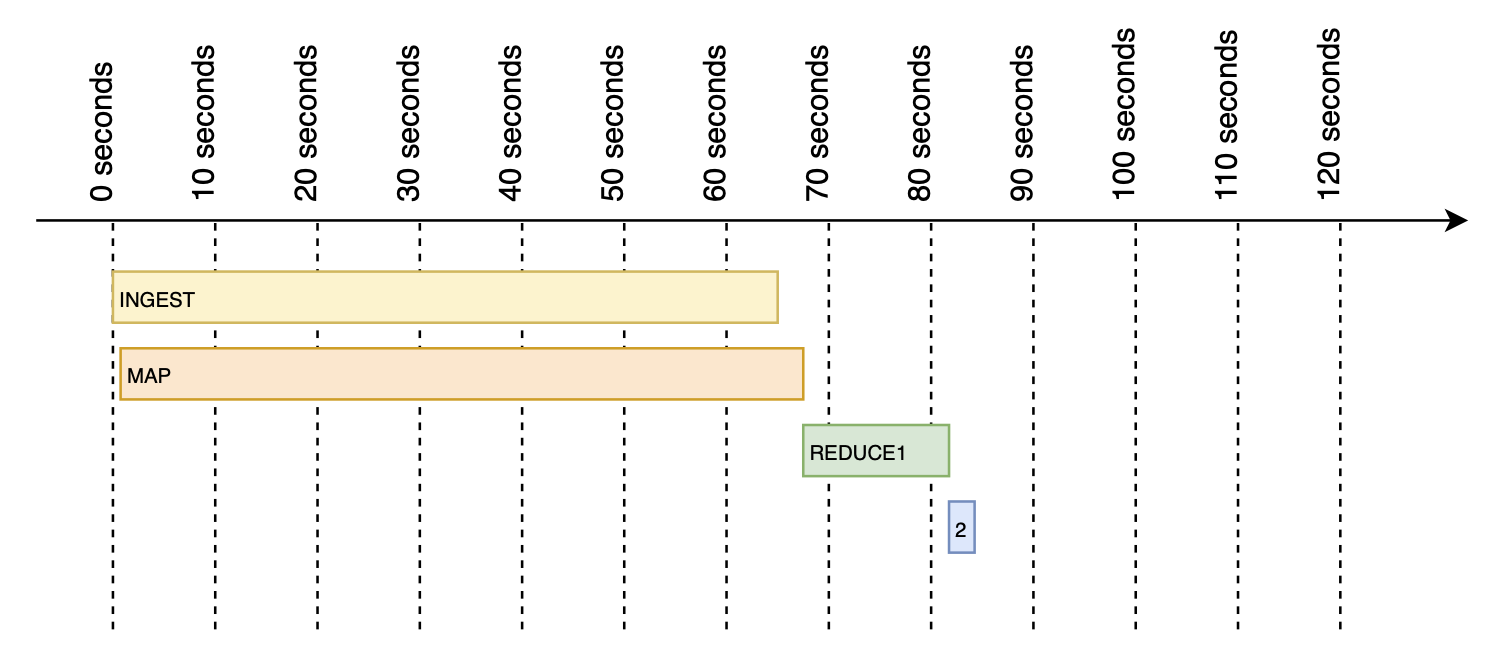}
\caption{2 threads for ingest, 2048/128/10240/128MB lambda memory}
\label{sc2}
\end{subfigure}

\begin{subfigure}{0.49\textwidth}
\centering
\includegraphics[width=\linewidth]{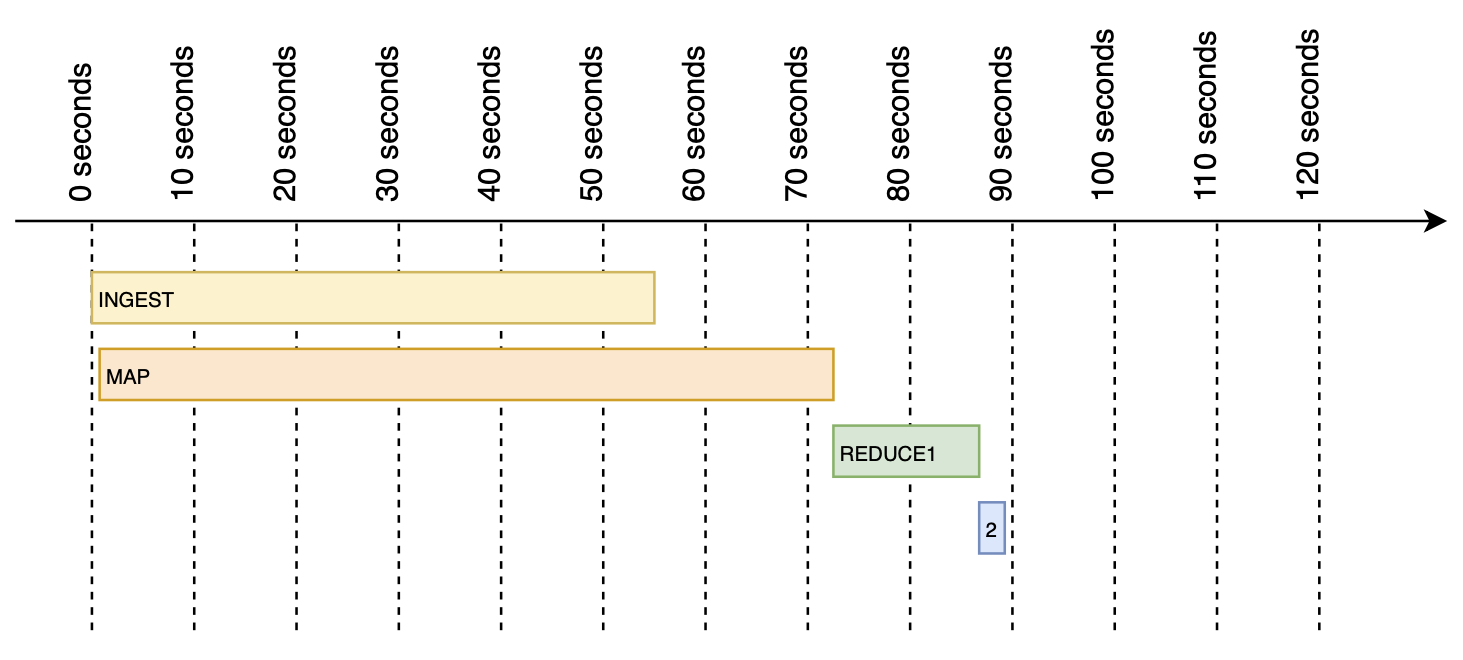}
\caption{3 threads for ingest, 3072/128/10240/128MB lambda memory}
\label{sc3}
\end{subfigure}
\begin{subfigure}{0.49\textwidth}
\centering
\includegraphics[width=\linewidth]{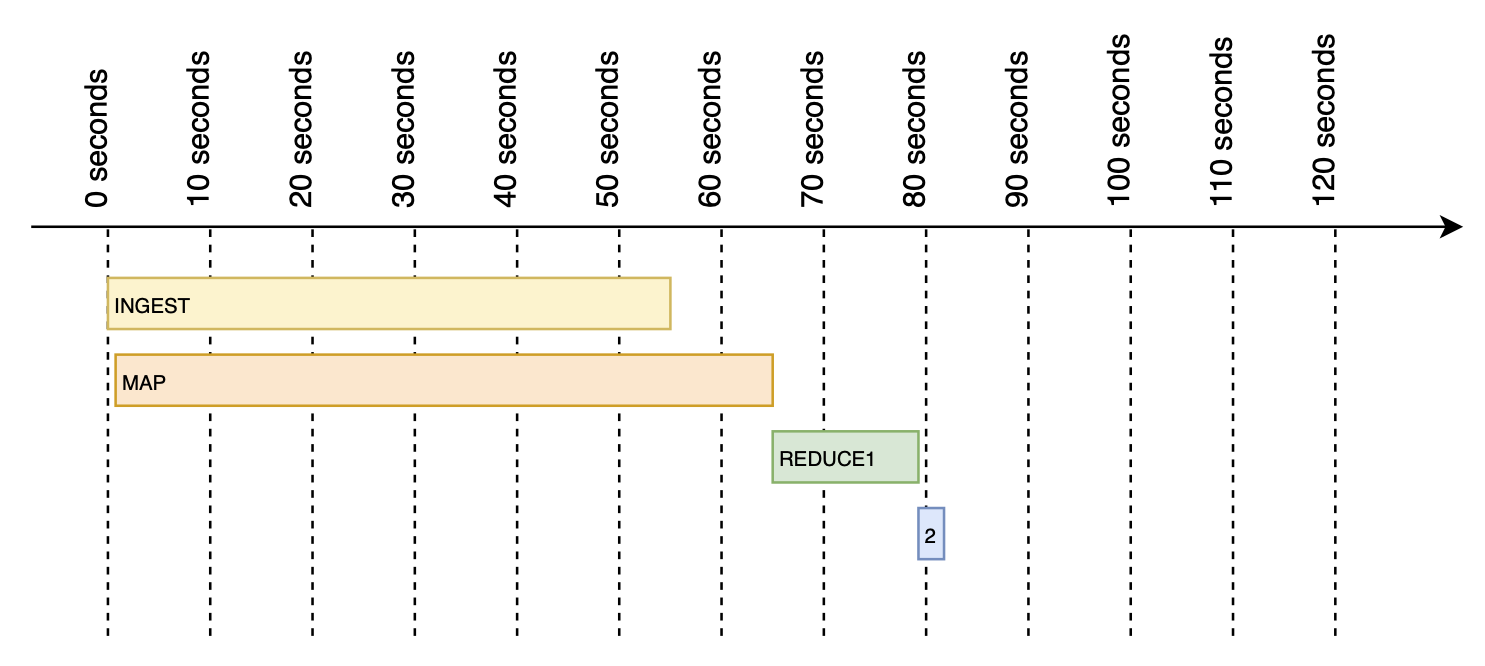}
\caption{3 threads for ingest, 3072/1024/10240/128MB lambda memory}
\label{sc4}
\end{subfigure}
\caption{S3 shuffle storage performance with 1 input file}
\label{s3shufflescenarios}
\end{figure*}

\subsection{Map}

For the map phase, we observed that writing individual S3 objects for each row in the CSV files introduced too much overhead for the read/write process. Instead, each micro-batch of records was grouped by partition key and written per microbatch. For example, if a microbatch had 100 total rows with 30 rows attributed to A, 30 rows attributed to B, 30 rows attributed to C, and 10 rows attributed to D, then four outputs result from a single map invocation, i.e. one for each partition key A, B, C, D. This reduced the overall latency associated with having to write that many objects or items and then later having to read them in the reduce phase. For example, a single 135MB CSV file had roughly 460k rows averaging 300 bytes per object. The 100x reduction reduced the object payload down to 4600 objects.

In our implementation, we used hexagonal architecture \cite{b40} when implementing the shuffle storage for the map and reduce functions. We did this by implementing an adapter and port pattern for S3 and DynamoDB, and then used configuration parameters via environment variables to determine which shuffle storage would be used. This allowed us to switch the underlying implementation from S3 to DynamoDB fairly easily when conducting performance tests.

One key consideration when analyzing the map function is that the map function begins executing almost as soon as the ingest function emits messages into SQS. So the key question becomes whether the map function keeps up with the rate of ingest, or if it falls behind over time. Each map invocation handles 100 records, performs some sorting, and then outputs the shuffled data to S3 or DynamoDB. It also tracks the number of records processed in a DynamoDB counter for the reduce gate later.

Fig.\ref{s3shufflescenarios} are visualizations of the results of the four different tests conducted in S3.

After completing tests with S3 as shuffle storage, we switched the shuffle storage to leverage an on-demand DynamoDB table. When ingesting 12 raw data files into DynamoDB, we observed that of the 5,202,096 records, we lost 312,507 records (6.0\%) due to throttling from DynamoDB and messages subsequently being sent to the configured dead letter queue. We recognize that we can provision read/write capacity units to the DynamoDB table to handle the load, but we will leave that as next steps for this research, to determine the appropriate capacity and the cost implications for needing that type of burst in the table.

Because the average duration of the map requests was lower compared to the ingest function, we also observed that the map functions had less frequent cold starts (13.5\% compared to ingest). However, as a percentage of overall duration, we observed that for the functions that did experience cold start, the impact of initialization was far higher ($<$1.5\% for ingest, \raisebox{0.5ex}{\texttildelow}30\% for map).

\subsection{Reduce Gate}

For the reduce phase, we implemented a counter gate before beginning the reduce phase.

\begin{lstlisting}[linewidth=\columnwidth,language=json,breaklines=true]
{
 "id": "c3f72186-2466-4daa-ae81-beb3f80a305e",
 "ingested": 1299481,
 "mapped": 1299481
}
\end{lstlisting}

The reduce gate is important to ensure that all map executions are completed prior to beginning the reduce aggregations and ranking. As stated in the architecture overview (Section \ref{archoverview}), the reduce gate will check to ensure that for execution id of this job the ingested counter and mapped counter are equal. We recognize that there could potentially be a race condition such that the ingest count could equal the mapped count before the map phase is truly completed. To handle this, the ingest phase does not write its values until after it has completed, while the map phase writes values as it performs its micro-batch processing.

We also observed in rare failure conditions with S3 and in throttling scenarios with DynamoDB that the ingest and mapped counters were never equivalent and required a manual override to allow the gate to pass.

In future research, we would like to consider implementing proper dead letter queue processing for situations that the map function happens to initially fail.

\begin{table*}[t]
\caption{Performance data of S3 as shuffle storage: Overall execution time}
\begin{center}
\begin{tabular}{cccccccc}
\textbf{Scenario}&\textbf{Ingest (s)} & \textbf{ReducePrep (s)} & \textbf{ReduceGate (s)} & \textbf{ReduceAggregate (s)} & \textbf{ReduceRank (s)} & \textbf{Overhead (s)} & \textbf{Total (s)}\\
\hline
1 & 92.19 & 0.84 & 1.13 & 14.13 & 0.98 & 0.37 & 109.64\\
2 & 63.10 & 0.82 & 1.71 & 13.27 & 0.89 & 0.47 & 80.25 \\
3 & 54.01 & 0.08 & 20.50 & 12.60 & 0.08 & 0.49 & 87.77\\
4 & 55.40 & 0.40 & 11.20 & 13.27 & 0.49 & 0.39 & 81.15\\
5 & 58.57 & 0.50 & 62.02 & 162.68 & 0.51 & 0.40 & 284.70\\
\hline
\end{tabular}
\label{table:s3shuffle}
\end{center}
\end{table*}

\begin{table*}[htbp]
\caption{Performance data of S3 as shuffle storage: Percentage of overall execution time per stage}
\begin{center}
\begin{tabular}{ccccccc}
\textbf{Scenario}&\textbf{Ingest (\%)} & \textbf{ReducePrep (\%)} & \textbf{ReduceGate (\%)} & \textbf{ReduceAggregate (\%)} & \textbf{ReduceRank (\%)} & \textbf{Overhead (\%)}\\
\hline
1 & 84.1 & 0.8 & 1.0 & 12.9 & 0.9 & 0.1\\
2 & 78.7 & 1.0 & 2.1 & 16.6 & 1.1 & 0.6 \\
3 & 63.5 & 0.1 & 21.0 & 14.8 & 0.1 & 0.6\\
4 & 70.0 & 0.5 & 11.6 & 16.7 & 0.6 & 0.5\\
5 & 20.6 & 0.2 & 21.8 & 57.1 & 0.2 & 0.1\\
\hline
\end{tabular}
\label{table:s3shuffle2}
\end{center}
\end{table*}

\subsection{Reduce Aggregation}
The first part of the reduce phase was the reduction per airline code. This entails a function invocation reading all the mapped outputs associated with each airline code. With S3, the reduce1 function would read all of the objects in a prefix associated with one airline with a prefix in the form of \textit{[execution\_id]/[airline\_code]/[instance\_id].json}, e.g. \textit{07b8c1d8-4e89-4969-83f3-72df580132f9/AA/01a63ff8-88e0-4ab9-aa0b-1451c5cb6f0e.json}. 

With DynamoDB, the reduce1 function would read all of the objects from a DynamoDB associated with the same structure. To accomplish this, we needed to create a DynamoDB table with execution id as the hash key, instance id as the sort key, and then leverage a local secondary index with the execution id as the hash key and now the airline key as the sort key. This allowed us to perform queries against the execution id and airline key while also maintaining individual records for each instance id. We considered creating composite keys, e.g. [execution\_id]/[airline\_code], but this created additional complexity in the query logic.

We observed that as the input files scaled up, there was a decent amount of variability, e.g. in the case of 60 input data files, AA had 36,434 mapped outputs, HP had 10,178 mapped outputs, and PS had only 422 mapped outputs. Because of this, we observed significant variance in the execution of this function, e.g. the faster ones took 14-25 seconds, while in some cases, a small handful of functions actually timed out with the 15 minute timeout.

\subsection{Reduce Ranking}
The second part of the reduce phase is to sort the first reduction aggregations and to list out the top 10 airlines by on-time arrival performance. This function executes fairly quickly, as it simply iterates over an array and sorts the results. We observed this function averaged 61 milliseconds.

\subsection{End-to-End}

Starting with S3 as shuffle storage, we tested a number of performance configurations and tweaked the memory setting of each Lambda function and the amount of parallel processing to incrementally improve performance.

We created a number of configuration scenarios as originally outlined at the beginning of the document in Table \ref{table:params}. Performance data for S3 as shuffle storage is outlined in Table \ref{table:s3shuffle}. The performance for DynamoDB is not included in this analysis due to significant throttling.

Of particular interest was that as more data was pushed into the system, the ingest phase became less of a bottleneck, as map started experiencing additional pressure. Even more, the reduce1 function (ReduceAggregate) in some cases needed to read a significant amount of mapped data. As a result, this phase began to take a larger percentage of the overall job execution as shown in Table \ref{table:s3shuffle2}.

\section{Conclusion}

Having implemented a prototype of a serverless map/reduce process that includes an ingest, map, and reduce phase, our observation is that the compute portion using AWS Lambda was able to scale up seamlessly and quickly, which is excellent for variable workloads and unpredictable execution times. This helps reduce the overhead of operating a long-lived cluster and eliminates the cost of idle infrastructure. However, our observation is also that the overhead of converting large data files into a message-oriented data flow and writing data back and forth between S3 or DynamoDB was substantial. The latter issue is a well-known issue in the Hadoop framework, hence the emergence of Spark processing in memory. The former issue is a new issue introduced by this serverless map/reduce framework which adopts an event-driven architecture for data processing. Because of this, we recognize that with the current implementation, large scale, long-running batch processing workloads would not likely be suitable on this from a cost and end-to-end completion time perspective. Enhancements will be required on the ingestion side and in the shuffle mechanisms to enable larger scale data processing workloads with this new serverless map/reduce framework.

\section{Future Work}
To improve overall performance, one potential opportunity would be to implement a third storage adapter for a caching layer. Concretely, we could perhaps use something like Elasticache Redis to load all of the data and perform all read/write operations across Lambda function invocations via this caching layer. This would more closely emulate the move from Hadoop using HDFS for shuffle storage to Spark doing all shuffling in memory.

For the ingest phase, we will need to further investigate methods of better parallelizing the workload, perhaps switching language runtimes from Python to a compiled language that would have better multi-threading support. We could also investigate switching from using SQS to a different streaming mechanism like Kinesis or Kafka. This could potentially allow for better Lambda consumer scaling, as SQS is limited on scaling up Lambda concurrency, e.g. 60 concurrency per minute if the queues are full.

For the reduce aggregation phase, we would need to determine a better way to handle the potential imbalance of processing the airline codes, which might include some scenarios that require far more processing than what a single Lambda function invocation can handle. If the aforementioned caching solution does not significantly improve performance for this aggregation phase, we can perhaps consider splitting the workload across multiple invocations, using Step Functions again to orchestrate the workflow.

\end{document}